\documentclass[graybox]{svmult}

\usepackage{mathptmx}       
\usepackage{helvet}         
\usepackage{courier}        
\usepackage{type1cm}        
\usepackage{makeidx}         
\usepackage{graphicx}        
\usepackage{multicol}        
\usepackage[bottom]{footmisc}
\usepackage{amsmath,amssymb}

\usepackage[bottom]{footmisc}

\usepackage{color}
\definecolor{keywordcolor}{rgb}{0,0,1}
\providecommand{\keyword}[1]{\index{#1}{\color{keywordcolor}#1}}

\makeindex             

\begin{document}

\title*{Free space interference experiments with single photons and single ions}
\author{Luk\'a\v{s} Slodi\v{c}ka, Gabriel H\'etet, Markus Hennrich, and Rainer Blatt}
\institute{Luk\'a\v{s} Slodi\v{c}ka \at Department of Optics, Palack\'y University, 17. listopadu 1192/12, 771 46 Olomouc, Czech Republic, \email{slodicka@optics.upol.cz}
\and Gabriel H\'etet \at Laboratoire Pierre Aigrain, Ecole Normale Sup\'erieure-PSL Research University, CNRS, Universit\'e Pierre et Marie Curie-Sorbonne Universit\'es, Universit\'e Paris Diderot-Sorbonne Paris Cit\'e, 24 rue Lhomond, 75231 Paris Cedex 05, France, \email{gabriel.hetet@lpa.ens.fr}
\and Markus Hennrich \at University of Innsbruck, Technikerstr. 25, 6020 Innsbruck, Austria, \email{markus.hennrich@uibk.ac.at}
\and Rainer Blatt \at University of Innsbruck, Technikerstr. 25, 6020 Innsbruck, Austria, and
Institute for Quantum Optics and Quantum Information, Austrian Academy of Sciences, Technikerstr. 21a, 6020 Innsbruck, Austria, \email{rainer.blatt@uibk.ac.at}}
%
%
\maketitle


\abstract{Trapped ion crystals have proved to be one of the most viable physical implementations of quantum registers and a promising candidate for a scalable realization of quantum networks. The latter will require the development of an efficient interface between trapped ions and photons. We describe two research directions that are currently investigated to realize such photonic quantum interfaces in free space using high numerical aperture optics. The first approach investigates how strong focusing of light onto a single ion can increase the interaction strength to achieve efficient interaction between a photon and the ion. The second approach uses a probabilistic measurement on scattered photons to generate entanglement between two ions that could be used to distribute information in a quantum network. For both approaches a higher numerical aperture would increase the efficiency of the interface.
\newline\indent
}

\section{Coupling to a single ion in free space} \label{sec:1}

Atom-photon interfaces are a key element for constructing a quantum network \cite{Cir97,Bri98,Dua01}. Here, the interface maps quantum information from stationary to flying qubits and vice versa. Usually photons are employed as flying qubits due to their robustness in preserving quantum information during propagation, while atoms are used for storing and computing the quantum information in stationary nodes. The efficient mapping of quantum information from atoms to photons and back demands controlled photon emission and absorption with a very high probability. This condition can be achieved in a strong coupling regime where the information is exchanged between atoms and photons several times before it decoheres. The standard way to achieve strong atom-photon coupling is by using either small high finesse cavities, which increase the interaction between a single atom and a photon \cite{Bru94,Pin00,Hoo00} as described in chapter by A. Kuhn, or large atomic ensembles for continuous variable quantum interfaces \cite{Pol04,Hau99,Phi01} treated in chapter by Chuu \& Du. 

In free space, i.e. without an enhancing cavity, the coupling of a single atom and light is generally considered to be weak. Nevertheless, it can be significantly increased if the light covers a large solid angle, for instance by using large aperture lenses \cite{Sor07} or mirrors \cite{Son07}. In such a setup it is possible to observe effects where a single atom can notably modify the light field. For instance, several experiments have recently demonstrated that a single quantum particle, like a single rubidium atom \cite{Tey08}, a molecule \cite{Zum08,Ger07,Wri08}, or a quantum dot \cite{Vam07} can cause extinction of more than 10\,\%, and a phase shift of one degree \cite{Alj09} for the transmitted light. Also, a single molecule has been shown to act as a nonlinear switch \cite{Hwa09}. These experiments are first steps towards realizing photonic quantum gates and quantum memories with single atoms in free space.

In this chapter we will discuss several experiments in which single ions are placed at the focus of \keyword{high numerical aperture optics} for efficient atom-light interaction. In section \ref{sec:1} of this chapter we will review several experiments on direct \keyword{free space coupling} of a \keyword{single trapped ion} with light. Here, the high numerical aperture optics allow the observation of effects which are usually only observed with large atomic ensembles or single atoms in high finesse cavities, including electromagnetically induced transparency \cite{Slo10}, coherent back scattering \cite{Het11}, and Faraday rotation induced on a propagating laser field \cite{Het13}. Very similar ideas and experiments discussing the efficient absorption of single photons by single ions are described also in the chapters by Leuchs \& Sondermann and by Piro \& Eschner.

Later in section \ref{sec:2} we will treat probabilistic methods to exchange quantum information over a distance. Here, a projective measurement on the fluorescence photons scattered by two atoms projects them into an entangled state, which then for instance could be used to link distant registers in a quantum network using protocols like quantum teleportation or entanglement swapping.

\subsection{Electromagnetically induced transparency from a single atom in free space}

The controlled storage and retrieval of photonic quantum information from an atomic medium is often based on a phenomenon called \keyword{electromagnetically induced transparency} (EIT)~\cite{Fle05} or its excitation in the form of stimulated Raman adiabatic passage (STIRAP)~\cite{Ber98a}. This technique has been widely used to control the storage of weak light pulses or single photons in atomic ensembles \cite{Phi01,Hau99,Eis05} and high-finesse cavities \cite{Boo07,Mue10}. For EIT, atoms in a lambda-type three-level system are driven by a weak probe laser and a strong control laser in Raman configuration. Due to a destructive quantum interference effect the control laser suppresses the absorption of the resonant probe light. Consequently, by changing the control laser intensity it is possible to switch the medium between transmitting and absorbing the probe light. Seen from a different point of view, the control laser intensity changes the group velocity of the probe laser. Thus, adiabatic ramping of the control laser intensity can slow down and even stop a single probe photon. In this way the photon is stored in the long-lived atomic ground states of the medium and can be retrieved by a time-reversal of the storage process \cite{Cir97,Fle05}. An extension of this scheme can store a photonic quantum state, for instance encoded in the polarization of light, in superpositions of atomic ground states \cite{Rit12}, thus realizing a memory for the photonic quantum information.

Effective switching between transmission and absorption can only be achieved in optically thick media. Therefore, until recently the application of EIT was restricted to ensembles of many atoms \cite{Fle05}. In contrast, (discrete variable) quantum information processing is based on single well-defined qubits (for example single atoms or ions) where each qubit can be individually manipulated to perform quantum gates. A quantum network which combines these two technologies requires strong single atom-single photon interaction within the interface nodes to distribute quantum information over the nodes of the quantum network.

Trapped ions are at the moment one of the most advanced systems for quantum information processing \cite{Hae08}. Also, since ions of the same species are identical, they are very well suited as indistinguishable light sources \cite{Mau07,Ger09njp} at the distant locations of a quantum network. Furthermore, the precise control over the electronic and motional states of the ions in Paul traps makes them ideal to investigate the coupling of radiation to single absorbers. In the following, we will present first steps towards a free-space single ion quantum interface by demonstrating an extinction of a weak probe laser of 1.3\,\%, electromagnetically induced transparency from a single trapped ion, and the corresponding phase shift response.

\subsubsection{Extinction and phase shift measurements}

\begin{figure}[t]
\begin{center}
\includegraphics[width=0.6\columnwidth]{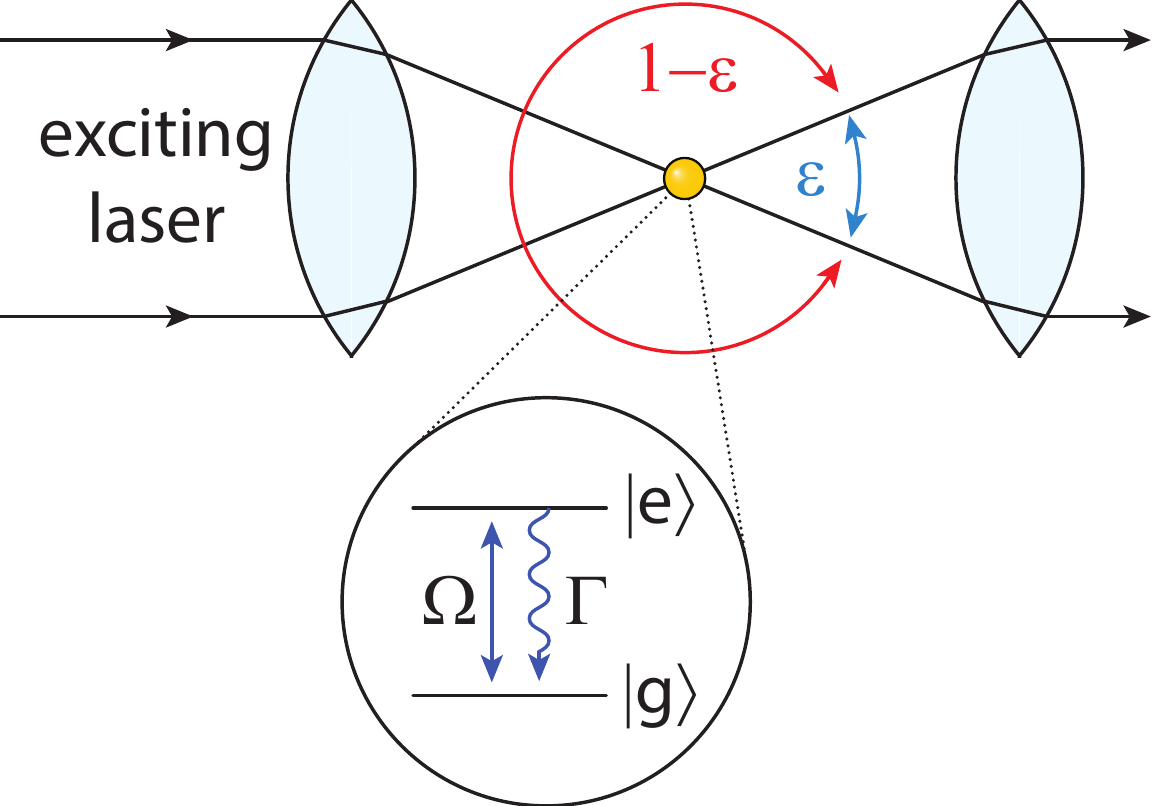}
\end{center}
\caption{Schematic view of a single atom irradiated by continuous laser light. The atom is illuminated from a fraction of the solid angle $\epsilon$ and radiates into full solid angle. The fluorescent light in general contains both elastic and inelastic components, nevertheless, for weak excitation the elastic component dominates. In the forward direction the elastically scattered part of the fluorescent light interferes with the transmitted laser beam ($\epsilon$). In the backward direction ($1-\epsilon$) the fluorescent light is observed.}
\label{Fig:2Lscheme}
\end{figure}

The experiments that we will describe in this chapter show the relation between the input and the transmitted light field in the presence of an atom \cite{Het13}. The following simple theoretical model can describe the basic properties of the \keyword{extinction} and reflection of a weak probe field from a single atom. This approach uses a perturbative \keyword{input-output formalism} to relate the input field, $\hat{E}_{\rm in}$, and the output field, $\hat{E}_{\rm out}$, through their interaction with the atom \cite{Koc94} as schematically depicted in Fig.~\ref{Fig:2Lscheme}. In Markov approximation, the output field in forward direction can be described as a superposition of the transmitted input field and the emitted field of the radiating atomic dipole as
\begin{equation}\label{EqInpOutp}
    \hat{E}_{\rm out}(t)=\hat{E}_{\rm in}(t) + i\sqrt{2\gamma_{\rm in}}\hat{\sigma}(t),
 \end{equation}
where $\hat{\sigma}(t)$ is the atomic coherence and $\gamma_{\rm in}$ is an effective coupling coefficient of the input field to the atom. The coupling coefficient can also be expressed by the total decay rate of the excited state $\gamma$ and the fraction $\epsilon$ of the full solid angle covered by the incoming field as $\gamma_{\rm in}=\epsilon\gamma$.
The atom reacts on the excitation by the input field, thus the atomic coherence $\hat{\sigma}(t)$ can be calculated by solving Bloch equations of the two-level atom in the weak excitation limit and in steady state, which gives
\begin{equation}
\hat{\sigma}=\frac{i \sqrt{2\gamma_{\rm in}}}{\gamma+i\Delta} \hat{E}_{\rm in},
\end{equation}
where $\Delta$ is the frequency detuning of the probe light from the excited state. Finally, the transmission of the intensity of the probe field $T=|E_{\rm out}/E_{\rm in}|^2$ in steady state reads
\begin{eqnarray}\label{Eqabs}
T(\Delta)=|1-2\epsilon\mathcal{L}(\Delta)|^2,
\end{eqnarray}
where the amplitude of the atomic coherence is proportional to $\mathcal{L}(\Delta)=\gamma/(\gamma+i\Delta)$ for a two level atom. Furthermore, the \keyword{phase shift} $\phi$ of the transmitted light field is
\begin{eqnarray}\label{ps}
\phi(\Delta)=\arg\big[1-2\epsilon \mathcal{L}(\Delta)\big].
\end{eqnarray}
Please note that this simple theory predicts perfect extinction of the transmitted probe field, and thus full reflection for a weak resonant input field covering half of the full solid angle ($\epsilon=0.5$). This results from the interference between the transmitted input beam and the radiated dipole field, which yields a considerable decrease in the forward mode amplitude \cite{Tey09,Zum08}. In the experiment described below, we use a lens with numerical aperture NA=0.4 (i.e $\epsilon=4\,\%$), so we expect a probe beam extinction of $16\,\%$ from Eq.~(\ref{Eqabs}) from this basic theoretical model. More refined models \cite{Venk04,Tey08,Zum08} include effects like the polarization of the input beam and the exact mode overlap between the transmitted beam and the dipole emission pattern which becomes especially important when the input beam is focused to the atom from a large solid angle with a high numerical aperture lens beyond the paraxial approximation. From the model of \cite{Tey08}, we expect an extinction of around $13\,\%$ for our experimental parameters. A higher extinction is possible for larger solid angles. Nevertheless, if one wants to use a single atom in free space as a quantum interface, for an efficient sender one will need to collect the emitted photon from full solid angle. Also, an efficient quantum receiver will demand reversal of the emission process, thus the single photon input mode will have to match the full dipole radiation pattern and the reversed temporal mode of the atomic emission as detailed in \cite{Son07} and chapter by Leuchs \& Sondermann.

\begin{figure}[ht!]
\centerline{\includegraphics[width=0.9\columnwidth]{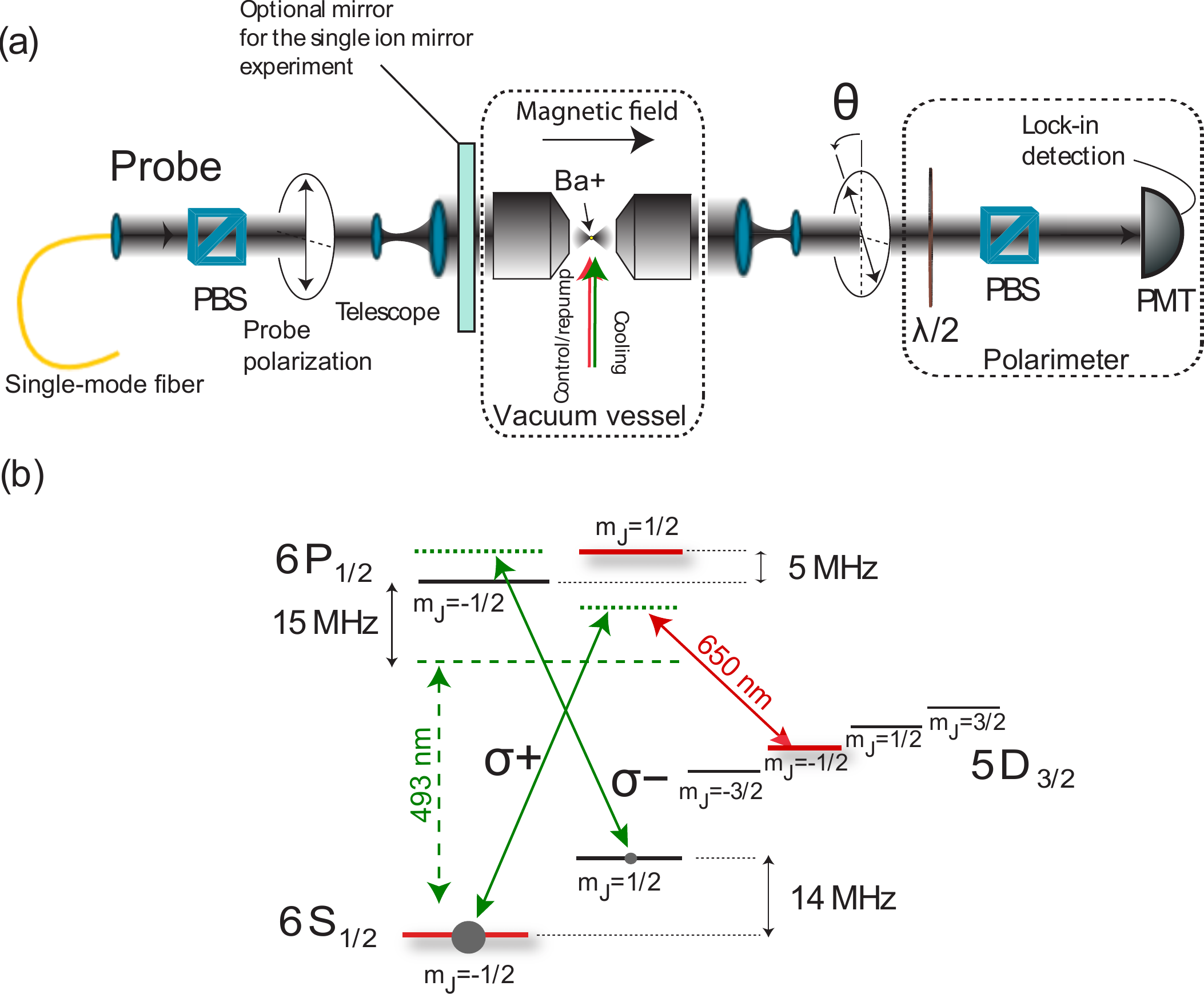}}
\caption{a) Scheme of the experimental set-up used to measure EIT and the corresponding phase shift on a single atom~\cite{Het13}. The probe field is defined in horizontal polarization by passing through a polarizing beam-splitter (PBS). The laser beam is then expanded by a telescope and focused by a high numerical aperture lens in vacuum (NA=0.4) onto the ion. Depending on the intensity of the control laser, the single ion changes the transmission of the probe light and rotates its polarization, which after re-collimation is detected by polarimetry.
b) Level scheme of $^{138}$Ba$^+$ and probe and control laser fields used in the experiment. The input probe field at 493\,nm is decomposed in the two circular polarizations which excite two branches of the spin-half system with different detunings. The laser field at 650\,nm is used as the control field in the EIT measurements and for repumping population from the $5D_{3/2}$ level.}
\label{setup}
\end{figure}

In the following we will describe how we measure the single ion transmission, the phase shift, and the EIT effect in our experiment. The experimental setup consists of a high numerical aperture objective \cite{Esc01} to focus the probe field onto a single trapped ion and a second objective to collect the transmitted light onto a detector, shown in Fig.~\ref{setup}-a). The barium ion is trapped and cooled in a standard spherical Paul trap. As already mentioned before, good extinction of the probe field can only be achieved if the incoming probe beam and dipole emission pattern are carefully overlapped. This mode-matching is done using an expanding telescope and a custom-designed objective with a numerical aperture of NA=0.4 ($\epsilon=4\,\%$).  A magnetic field of 5 Gauss applied along the probe beam propagation direction defines the quantization axis. The probe field is linearly polarized perpendicular to the quantization axis, collected after the ion using a second high numerical aperture lens, and then analyzed using a polarimetric set-up and photo-multiplier tubes (PMT). In practice, we detect the transmitted polarization alternately at +45 or -45 degrees with respect to the input polarization.

The level scheme of Ba$^{+}$ is shown in Fig.~\ref{setup}-b). The probe field is tuned to the $6S_{1/2}\rightarrow 6P_{1/2}$ transition and with our choice of quantization axis, its polarization can be decomposed onto left and right circularly polarized modes that drive $\sigma_-$ and $\sigma_+$ transitions in the ion, respectively. The two polarization modes do not have the same detuning from their respective transitions and thus may experience different indices of refraction. We set the intensity of the probe field well below saturation so that most of the light is elastically scattered. The probe field supplies only weak cooling to the ion. Therefore the ion is cooled additionally by a red detuned 493\,nm laser field perpendicular to the probe direction and a laser at 650\,nm, co-propagating with the cooling beam, for pumping out population from the $5D_{3/2}$ level. For the characterization of the Faraday rotation we use a lock-in method where the 650\,nm repumping laser is switched on and off at a rate of 5\,kHz to precisely measure the polarization rotation signal. When the repumping laser is off the population of the atom is pumped into the $5D_{3/2}$ state. Since the probe beam drives the transition between $6S_{1/2}$ and $6P_{1/2}$, it does not feel the presence of an ion that resides in $5D_{3/2}$. Thus, the modulation of the repumping laser also modulates the effect of the atom on the probe field. The photo-multiplier signal is demodulated and low-pass filtered with a time constant of 1s.

The intensity of the light at the PMT measured at +45 degrees with respect to the input polarization can be written as
\begin{eqnarray}
I_{45}=\frac{1}{2}|E^+_{\rm out} e^{i\pi/4}+E^-_{\rm out}  e^{-i\pi/4} |^2,
\label{eqn:pol45}
\end{eqnarray}
where
\begin{eqnarray}
E^+_{\rm out} &=(1-2\epsilon \mathcal{L}^+) \frac{1}{\sqrt{2}} E_{\rm in}&\quad\mathrm{and} \\
E^-_{\rm out} &=(1-2\epsilon \mathcal{L}^-) \frac{1}{\sqrt{2}} E_{\rm in}&
\end{eqnarray}
are the corresponding output fields of $\sigma_+$ and $\sigma_-$ polarization, respectively.  The real and imaginary parts of
\begin{equation}
\mathcal{L}^{\pm}=\frac{\rho_\pm\gamma}{\gamma+i\Delta^{\pm}}
\end{equation}
correspond to absorption and phase-lag of the two scattered circularly polarized field modes with regards to the input field, respectively. Here, $\Delta^{\pm}=\Delta \pm \Delta_{B}$ are the detunings of the $\sigma^+$ and $\sigma^-$ polarized fields from their respective transitions. The two ground state populations $\rho_+$ ($\rho_-$) of the $\sigma_+$ ($\sigma_-$) transition correspond to the states $6S_{1/2}$, $m_J=-1/2$ ($m_J=+1/2$), respectively. $\Delta$ is the probe laser detuning with respect to the $6S_{1/2} \rightarrow 6P_{1/2}$ transition without Zeeman shift, and $\Delta_{B}$ is the frequency offset of the two resonances due to the Zeeman splitting. The $\pm\pi/4$ phase shifts in Eq.~(\ref{eqn:pol45}) are due to the rotation of the polarization direction between input and detected polarization induced by the $\lambda/2$ waveplate that allows us to characterize the Faraday rotation.

To measure the Faraday rotation angle $\theta = \frac{1}{2} \arctan(s_2/s_1)$, here defined as half the rotation angle from horizontal polarization towards 45 degree polarization on the Poincar\'e sphere, we need to record the Stokes parameter $s_1=I_{0}-I_{90}$, and $s_2=2 I_{45}-s_0$, with $s_0=I_{0}+I_{90}$. Please note that for our small extinction values the Stokes parameter $s_1$ can be approximated by $s_1\approx s_0\approx I_{0}$, which can be measured directly by removing the polarizing beam-splitter. After re-inserting the beam-splitter and adjusting the waveplate accordingly, we can access the Stokes parameter $s_2\approx 2 I_{45}-I_0$. The Faraday rotation angle $\theta \approx \frac{1}{2} \arctan((2 I_{45}-I_0)/I_0)$ is directly related to the phase shift induced by the atom. It can be shown, using the approximation $\arg(1-2\epsilon z)\approx -2\epsilon {\rm Im}(z)$ in the limit of small $\epsilon$, that
\begin{eqnarray}\label{ps1}
\theta=\frac{1}{2}\arg\big[1-2\epsilon ( \mathcal{L}^+ - \mathcal{L}^- )\big],
\end{eqnarray}
which is half of the phase lag experienced by the output with respect to the input field. A measurement of $I_{45}$ and $I_{0}$ thus provides a measurement of the Faraday rotation of the light across the atom together with the phase difference acquired by the two circularly polarized modes.

\begin{figure}[!t!]
\hspace*{-2mm}\centerline{\scalebox{0.47}{\includegraphics{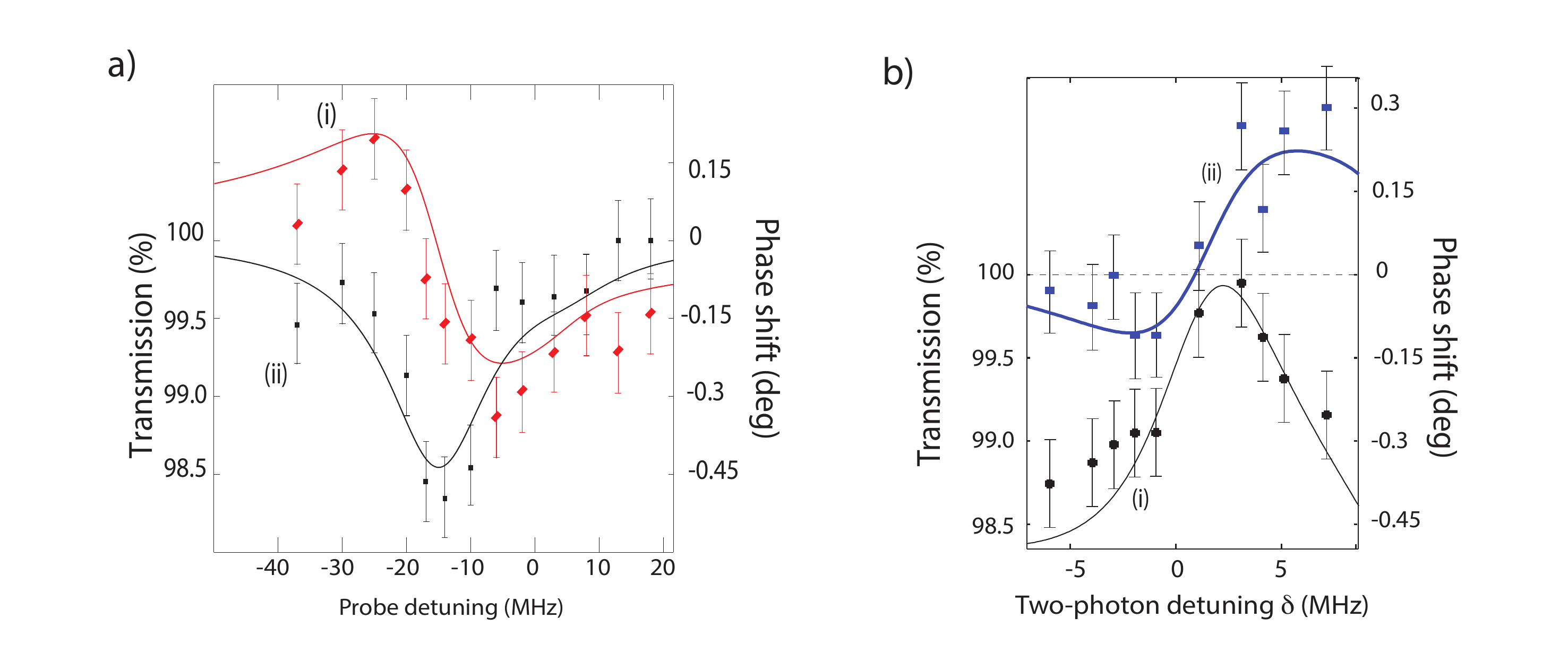}}}
\caption{a) Transmission $I_0$ (ii) and phase shift $\theta$ (i) of a probe field transmitted through a single trapped barium ion as a function of probe beam detuning~\cite{Het13}. The transmission spectrum is fitted by a Lorentzian profile with a width of 11\,MHz. The peak probe beam extinction is 1.35\,\%. b) Transmission (i) and phase shift (ii) of a probe field transmitted through a single trapped barium ion as a function of probe beam detuning close to a dark resonance.}
\label{absorption}
\end{figure}

We characterize the Faraday rotation of the probe field by measuring the phase shift $\theta$ and the transmission $I_0$, which are plotted in Fig.~\ref{absorption}-a) as a function of the probe frequency detuning $\Delta$. As can be seen from the measurement of $I_0$, an optical pumping mechanism by the cooling and repumping lasers causes a strong unbalancing between the two ground states populations. In principle, one would expect to observe two identical absorption lines at +5\,MHz and -14\,MHz (shifted with respect to the symmetric case due to dipole shifts) for the $\sigma_+$ and $\sigma_-$ transitions of the probe light from $6S_{1/2}$ to $6P_{1/2}$, respectively. Here, the optical pumping induced by the cooling and repumping lasers traps population in the $| 6 S_{1/2}, m_J=+1/2 \rangle$ level so that only the $\sigma_-$ transition can be detected by the probe field. This manifests itself in the 1.5\,\% extinction that is seen -14\,MHz red-detuned from the central line for the $\sigma^-$ and in the almost completely suppressed extinction for the other $\sigma^+$ mode at 5\,MHz. With this state preparation, trace (i) displays a clear dispersive profile across the resonance of the $\sigma_-$ transition and the $\sigma^+$ polarized mode is almost not phase shifted. Even with our small magnetic field, the pumping technique thus allows us to isolate a single two-level atom and to reach a maximum of 0.3 degrees phase-shift. Solid lines show the result of a fit of the data using the above four-levels calculations, with $\epsilon=0.8\,\%$, $\Delta_B=9$\,MHz, $\rho_-=0.9$ and $\rho_+=0.1$. With these parameters, good agreement is found with the experimental results.

\subsubsection{Electromagnetically induced transparency and associated phase shift with a single atom}

In the above measurements, the cooling and repumping beams were tuned to a dark resonance with the intention to pump the population of the ion into one of the $S_{1/2}$ levels and therefore minimize the population in the $D_{3/2}$ state which is not interacting with the probe laser. Nevertheless, the transverse cooling beam can also be turned off and the ion be cooled by the linearly polarized probe field itself.
In such a configuration the probe undergoes \keyword{electromagnetically induced transparency} (EIT)~\cite{Slo10} where the population in the excited state of the $\Lambda$ scheme (see Fig.~\ref{setup}-b)) is canceled due to a quantum interference between the two excitation pathways leading to the $P_{1/2}$ excited state.

Under weak probe excitation, the probe transmission as a function of the two-photon detuning $\delta=\Delta_g-\Delta_r$ can be found by solving the Bloch equations \cite{Ima89} and using the above input-output relations. We neglect here the angular dependence of the extinction (due to polarization). That is, we suppose that the probe has a polarization profile that matches the dipole field. This is a good approximation for the relatively small numerical aperture we use in this experiment. We can thus replace the function $\mathcal{L}$ by
\begin{eqnarray}\label{EITeq} \mathcal{L}_{\Lambda}(\delta)=\frac{\gamma(\gamma_0-i\delta)}{(\gamma_0-i\delta)(\gamma+i\Delta_g)+\Omega_r^2},
\end{eqnarray}
in Eq.~(\ref{Eqabs}), where $\Omega_r$ is the Rabi frequency of the red laser field, $\gamma_0$ the ground state dephasing rate, $\gamma$ the natural linewidth of the two transitions (assumed to be the same for simplicity). An important condition for EIT to take place is $\gamma\gamma_0 \ll \Omega_r^2$, i.e. the pumping rate to the dark state must be much faster than any ground state decoherence process. Independent frequency fluctuations of the two laser fields, magnetic field fluctuations, and atomic motion induced Doppler shifts, must be therefore reduced. When this is the case, extinction of the resonant probe can be completely inhibited, within a small range of control laser detuning $\Omega_r^2/\gamma$, creating an EIT window. This is what we observed in this experiment.

In the experiment we found that the motion induced decoherence yields broadening of tens of kHz, which reduced the EIT when the control and the probe were orthogonal to each other. However, the effect of Doppler shifts due to the ion motion could be eliminated when we used co-propagating control and probe fields. Since for optimum EIT conditions we could not use the cooling fields which would have reduced the transparency achieved through EIT, so the ion needed to be cooled by the probe itself. Consequently, the probe beam was more intense and red detuned, which resulted in reduced extinction efficiencies of about $0.6\,\%$. Additionally, we have to note here that due to the multi-level structure of barium, a single three level system can only be perfectly isolated from the others through optical pre-pumping. Therefore, Stark-shifts induced by the other levels and double-$\Lambda$ type couplings contribute to a slight reduction of the EIT contrast.

The results of the measurement of the probe transmission versus the two-photon detuning $\delta$ are shown in Fig.~\ref{absorption}-b) trace (i). In this EIT regime, a rapid change of the transmission is found as a function of the two-photon detuning and an almost complete cancellation of the transmission is measured at $\delta=0$.

Associated with such a steep change of the probe transmission, we also expect a fast roll-off of the phase. Fig.~\ref{absorption}-b)-trace (ii) shows the measurement of the phase $\theta$ of the probe field, using the same polarimetric technique as in the measurement described in the previous section. Here again, close to the dark resonance, the Faraday rotation angle yields the phase-shift induced by the atom. The clear dispersive shape of $\theta$ across the two-photon resonance is here a sign of the EIT induced phase-shift from the ion where a maximum phase lag of 0.3 degrees is observed. The solid lines show a fit to the experimental results using 8-level Bloch equations, consisting of the 2 $S_{1/2}$, the 2 $P_{1/2}$ and the 4 $D_{3/2}$ states. Here we replace the two-level atom Lorentzian functions $\mathcal{L}^{\pm}$ in Eq.~\ref{ps} by the newly found susceptibilities. The theory describes well the data with the repumping and probe field intensities as the only two free parameters. The asymmetry of the dispersion and transmission profiles that we measured is due to a slight overlap with neighboring dark-resonances and our detuned driving of the $\Lambda$ scheme. The distinctive feature of this interference effect is that the flipping of the phase shift sign occurs only over a couple of MHz. Increasing the slope steepness further can in fact be done by performing the experiment with smaller probe and repumping powers which can be implemented by appropriate switching of the laser cooling beams involved the experiment. Achieving a very steep phase shift dependence across the atomic spectrum would open the way for reading out the motional and internal energy of the atom.

Tightly focusing a weak, detuned, linear polarized probe field onto a single barium ion thus enables observation of both the direct extinction of a weak probe field and electromagnetically induced transparency from a single barium ion.  Besides demonstrating further the potential of these effects for fundamental quantum optics and quantum information science, these experimental results will trigger interest for quantum feedback to the motional state of single atoms, as proposed in \cite{Rabl} using EIT, for dispersive read out of atomic qubits and for ultra-sensitive single atom magnetometery.

In the following we will now discuss another effect observed with a similar experimental apparatus where we show that a single atom can act as a mirror of an optical cavity.

\subsection{Single ion as a mirror of an optical cavity}

Atom-photon interactions are essential in our understanding of quantum mechanics. Besides the two processes of absorption and emission of photons, coupling of radiation to atoms raises a number of questions that are worth investigating for a deeper theoretical and thus interpretational insight. The modification of the vacuum by boundaries is amongst the most fundamental problems in quantum mechanics and is widely investigated experimentally. We here present the very first steps towards merging the field of cavity QED with free-space coupling, using an ion trap apparatus.

Here we report an experiment where we set up an atom-mirror system~\cite{Het11}. As shown Fig.~\ref{setup}-a), we place a mirror in the path of the probe beam in front of the ion. The idea of this geometry is to form a \keyword{atom-mirror cavity} system consisting of the normal mirror and the ion acting as the second mirror. We observe the modification of the probe transmission and reflection of this atom-mirror cavity. Here, the atomic coupling to the probe is modified by the single mirror in a regime where the probe intensity is already significantly altered by the atom without the mirror.  In principle, in the limit of an even higher numerical-aperture lens, the mirror-induced change in the vacuum-mode density around the single atom could modulate the atom's coupling to the probe, the total spontaneous decay and the Lamb shift, so that the atom would behave as the mirror of a high-finesse cavity.

As before, for extinction of a laser field by the ion in free space, we use a very weak probe beam resonant with the S$_{1/2}(m_J=+1/2)$-P$_{1/2}(m_J=-1/2)$ transition. In the case of coherent reflection of a laser field by a single atom, the backscattered field must interfere with the driving laser. To verify this, we construct the system shown in Fig.~\ref{setup} a) by inserting a dielectric mirror 30\,cm away from the atom into the probe path, with a reflectivity $|r|^2=1-|t|^2=25\,\%$. We align it so that the ion is re-imaged onto itself and shine the resonant probe through it. Using the Fabry-P\'erot cavity transmissivity, and modeling the atom as a mirror with amplitude reflectivity $2\epsilon$~\cite{Koc94}, one can naively assume that the intensity transmissivity of the probe reads
\begin{eqnarray}\label{EQ1}
T=\Big|\frac{t(1- 2\epsilon) }{1- 2r\epsilon e^{i\phi_L} }\Big|^2,
\end{eqnarray}
where $\phi_L=2k_LR$, $R$ is the atom-mirror distance and $k_L$ the input probe wavevector. The finesse $\mathcal{F}=\pi 2\epsilon r/(1-(2\epsilon r)^2)$ of such a cavity-like set-up can in fact be made very large by using a high numerical aperture lens such that $\epsilon\rightarrow 50\,\%$ together with a highly reflective dielectric mirror. By tuning the distance between the dielectric mirror and the ion, one would therefore expect a dependence of the transmitted signal on the cavity length, provided that the temporal coherence of the incoming field is preserved upon single-atom reflection.

\begin{figure}[t]
\centerline{\scalebox{0.2}{\includegraphics{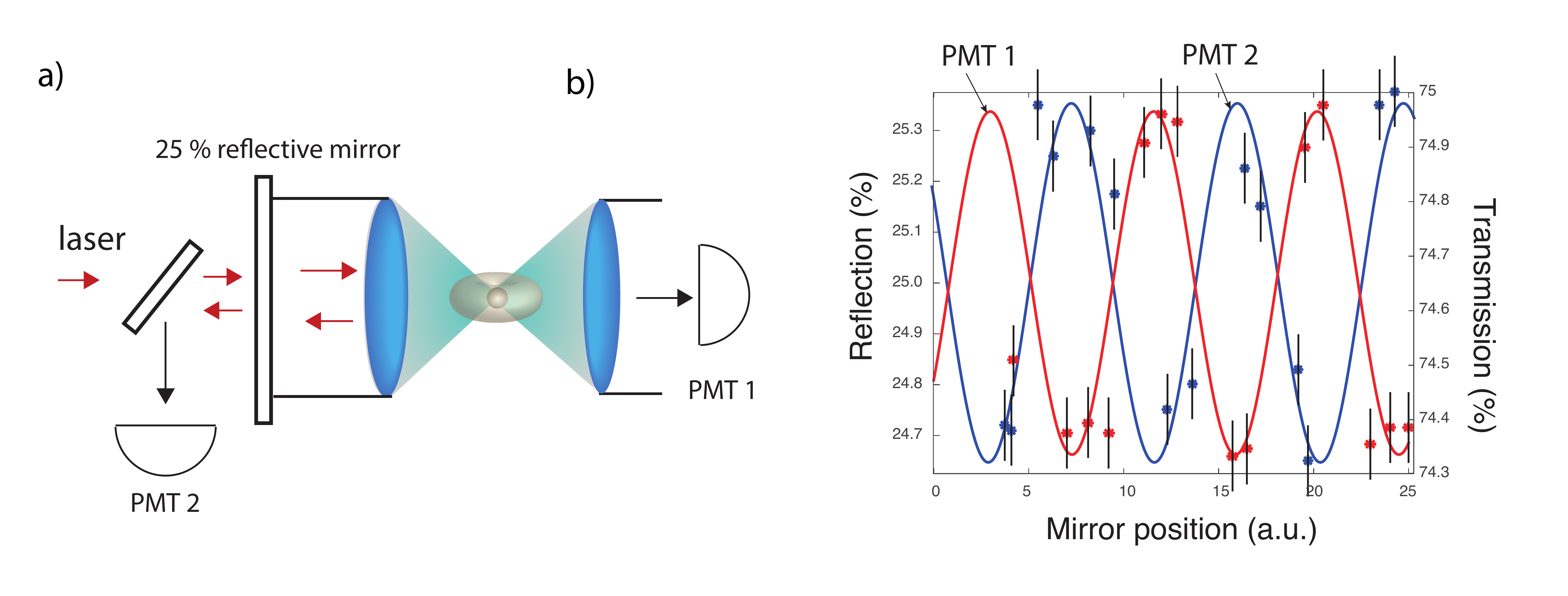}}}
\caption{a) Experimental setup with an optical cavity formed by the mirror and the single ion. PMT 1 and 2 measure the transmitted and the reflected probe laser intensity, respectively. b) Reflection and transmission signal of the ion-mirror cavity as a function of the mirror position~\cite{Het11}.
}\label{Fig:back}
\end{figure}

The operation of our ion-mirror system is shown in Fig.~\ref{Fig:back}~b), where we simultaneously recorded the reflected and transmitted intensity. As the mirror position is scanned, we indeed observed clear sinusoidal oscillations of the intensity on a wavelength scale. These results reveal that the elastic back-scattered field is interfering with the transmitted probe, and that the position of the ion is very well defined, meaning that it is well within the \keyword{Lamb-Dicke regime}. Reflected and transmitted intensity have opposite phase, as is predicted for a Fabry-P\'erot cavity response. The slight shift in the two sinusoidal fitting functions is within the measurement error bars.

We now investigate whether the naive Fabry-P\'erot interpretation that we used to describe our results is valid. One could indeed wonder how the modification of the quantum vacuum around the atom affects our results. It is clear that the dielectric mirror imposes new boundary conditions that change the \keyword{vacuum mode density} close to the atom, but it is less obvious how much this change contributes to the probe intensity modulation that we observe in this experiment. One can in fact show \cite{Het11} that solving the multimode Heisenberg equations in a time-dependent perturbation theory gives
 \begin{eqnarray}\label{EQ3}
T=|t|^2\Big|1-\frac{2g_{\epsilon}\overline{g}^\ast}{\tilde{\gamma}+i\tilde{\Delta}}\Big|^2,
\end{eqnarray}
assuming the input probe to be resonant with the atomic transition. Here, $g_{\epsilon}$ denotes the atomic coupling strength in the probe mode, $\overline{g}$ is the mean coupling to all the modes, $\tilde{\gamma}$ and $\tilde{\Delta}$ are the decay and level shifts modified by the presence of the mirror, respectively. Their value can be calculated using the appropriate spatial mode function for this system \cite{het10} and we can then show that
\begin{eqnarray}\label{EQ5}
\frac{g_{\epsilon}\overline{g}^\ast}{\tilde{\gamma}+i\tilde{\Delta}}= \frac{\epsilon(1- r e^{i\phi_L}) }{1- 2r\epsilon e^{i\phi_L} }.
\end{eqnarray}
After combining this relation with Eq.~\ref{EQ3} we obtain the same transmissivity as was obtained by modeling the atom as a mirror with reflectivity $2\epsilon$ (Eq.~\ref{EQ1}). Interestingly, the QED calculations yield the same mathematical results as the direct Fabry-P\'erot calculation.

In this \keyword{QED} approach, it was not necessary to invoke multiple reflections off the atom for the Fabry-P\'erot like transmission to appear. The transmission of the probe through the single atom+mirror system is mathematically equivalent to a cavity, therefore the origin of the peaked transmission profile can be interpreted either as a cavity effect or as a line-narrowing effect due to the QED-induced changes of the spontaneous emission rate and level shift. In the second interpretation, the observed oscillations can be interpreted as a change of the coupling between the atom and the probe mode, due to the modification of the mode density at the position of the ion induced by the mirror. For very high numerical optics the change of the extinction contrast would be analogous to an almost complete cancellation and enhancement by a factor of two of the atomic coupling constant in the probe mode. Deviations from the sinusoidal shape due to line narrowing would already be visible for a lens covering a solid angle of more than 10\,\%.

\section{Probabilistic entanglement between distant ions} \label{sec:2}

Long-lived \keyword{entanglement} between distant physical systems is an essential primitive for quantum communication networks~\cite{Bri98,Dua01}, and distributed quantum computation~\cite{Jia07,Cir99,Got99}. There are several protocols generating entanglement between distant matter qubits \cite{Dua10}, like single atoms. The majority of them exploit traveling light fields as mediators of the entanglement generation process. A way to generate distant entanglement is based on the spontaneous generation of entanglement between an atom and a single photon during the emission process followed by the absorption of the photonic state in a second atom~\cite{Cir97}. This method can generate entanglement deterministically, if the photon collection and absorption processes are highly efficient. Nevertheless, photon losses in experimental realizations might render it necessary to first detect a successful photon absorption in order to herald the successful entanglement generation. Another approach generates the entanglement probabilistically by detecting single photons that where scattered by two atoms. The projective measurement on the photons heralds an entangled state of the two atoms~\cite{Cab99, Sim03, Dua01, Bri98}.

The realization of heralded entanglement between distant atomic ensembles \cite{Cho05,Cho07} was amongst the first major experimental achievements in this field. Probabilistic generation of heralded entanglement between single atoms~\cite{Sim03} was demonstrated using single trapped ions~\cite{Moe07} with an entanglement generation rate given by the probability of coincident detection of two photons coming from the ions~\cite{Zip08,Luo09}. More recently, single neutral atoms trapped at distant locations were entangled using the deterministic entanglement protocol described above~\cite{Rit12}. Nevertheless, the efficiency of this realization was still limited due to losses to approximately 2\,\%. A heralding mechanism will therefore be essential for efficient entanglement and scalability of quantum networks using realistic channels \cite{Dua10}. The distant entanglement could also be used for one-way quantum computation schemes~\cite{Rau01,Dua05}. Such schemes for distributed quantum information processing would require only projective measurements and single qubit operations to perform quantum calculation~\cite{Dua10}. For future quantum information applications it therefore will be important to realize \emph{\keyword{heralded}} distant entanglement with the possibility of \emph{single qubit operations} and with \emph{high \keyword{entanglement generation rate}} at the same time.

\subsection{Single-photon and two-photon protocols}

The main limitation for generation of heralded distant entanglement between single atoms with high rate is imposed by relatively small overall detection efficiencies $\eta$ of fluorescence photons emitted by atoms trapped in free space \cite{Zip08}. For state-of-the-art experimental setups employing high numerical aperture optics close to single trapped neutral atoms or ions, $\eta$ is on the order of $10^{-3}$~\cite{Str11,Str12,Dar05,Olm10,Tey09}. There is a large effort in the experimental quantum optics community towards increasing this number both by employing very high numerical aperture optics in the form of spherical~\cite{Shu11} or parabolic~\cite{Mai09,Sto09} mirrors and by developing single-photon detectors with high quantum efficiency. However, even with these improvements it will be hard to increase the overall detection efficiency by more than one order of magnitude in the near future.

\begin{figure}[t]
\begin{center}
\includegraphics[scale=0.35]{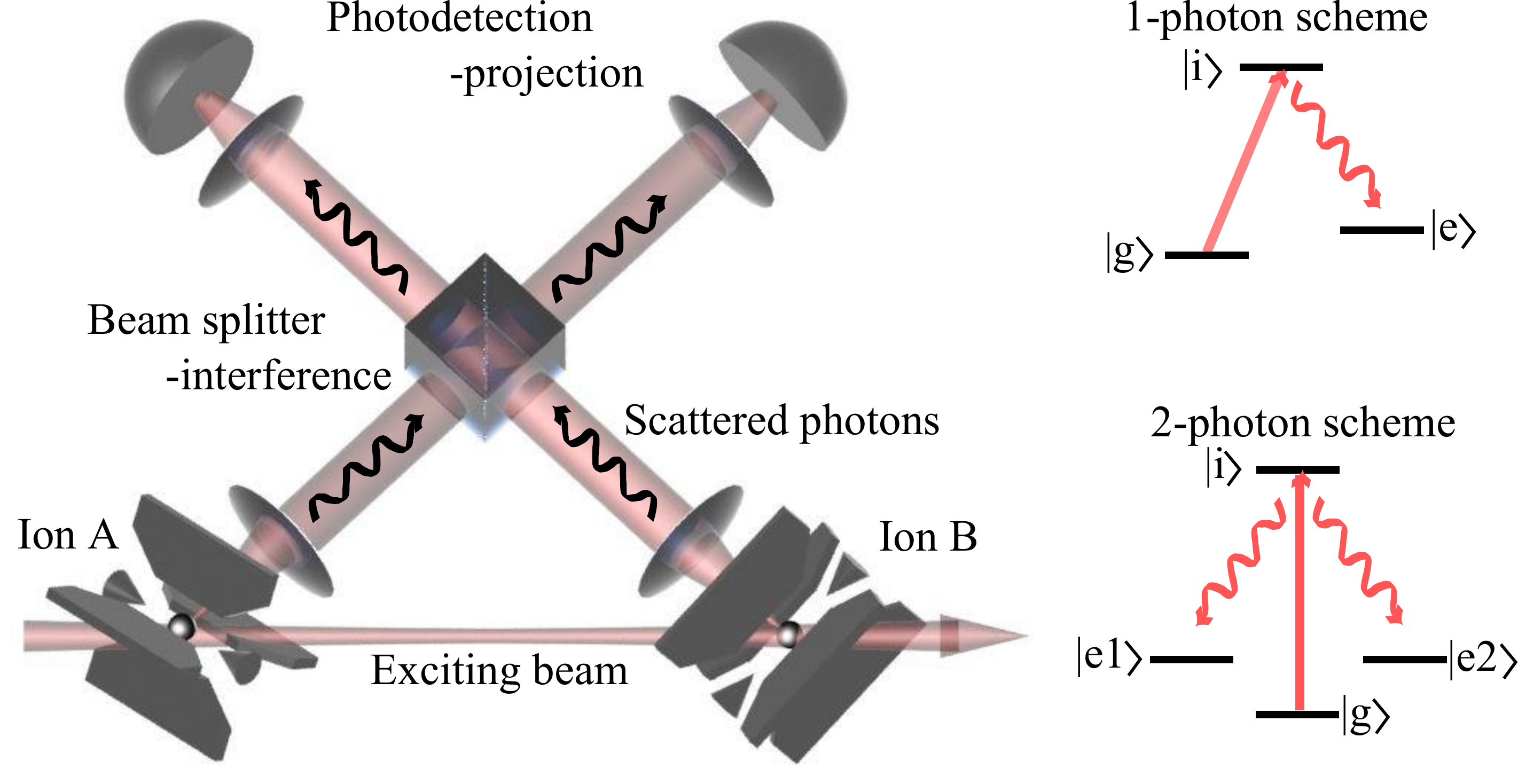}
\caption{Both one- and two-photon entanglement protocols use the same steps for generating heralded entanglement of distant atoms. After electronic excitation the atoms decay back to a ground state, while spontaneously emitting a photon. The scattered photons interfere to make it indistiguishable which of the two atoms has scattered the photon. Finally, the photons are detected and project the atoms into entangled state. For the two-photon protocol, two photons must be emitted from two distant atoms and the detection corresponds to the projection onto one of the Bell states in the photon basis. For the single photon scheme, only one photon has to be emitted and detected. Depicted energy levels correspond to the typical schemes employed for the two protocols with $|g\rangle$, $|i\rangle$ and $|e\rangle$ corresponding to the initial ground state, auxiliary excited state and final state after Raman process, respectively.}
\label{fig:protocolsEnt}
\end{center}
\end{figure}

We compare the efficiency of the two known heralded entanglement generation protocols based on the single-photon~\cite{Cab99} and two-photon~\cite{Sim03} detection. Both protocols are based on the atomic excitation, indistinguishability and \keyword{interference} of the emitted photons and on state-projective detection, as illustrated in Fig.~\ref{fig:protocolsEnt}.

We define two dimensionless measures crucial for the performance of any practical quantum information network~\cite{Zip08}. \keyword{Fidelity} between the generated state described by the density matrix $\rho$ and the desired maximally entangled two qubit state $|\psi\rangle$,
\begin{equation}
F = \langle \psi |\rho|\psi\rangle
\end{equation}
and success probability $P_{\rm s}$, corresponding to probability with which this state can be generated for given overall detection efficiency $\eta$.

\begin{figure}[!ht!]
\begin{center}
\includegraphics[scale=1.0]{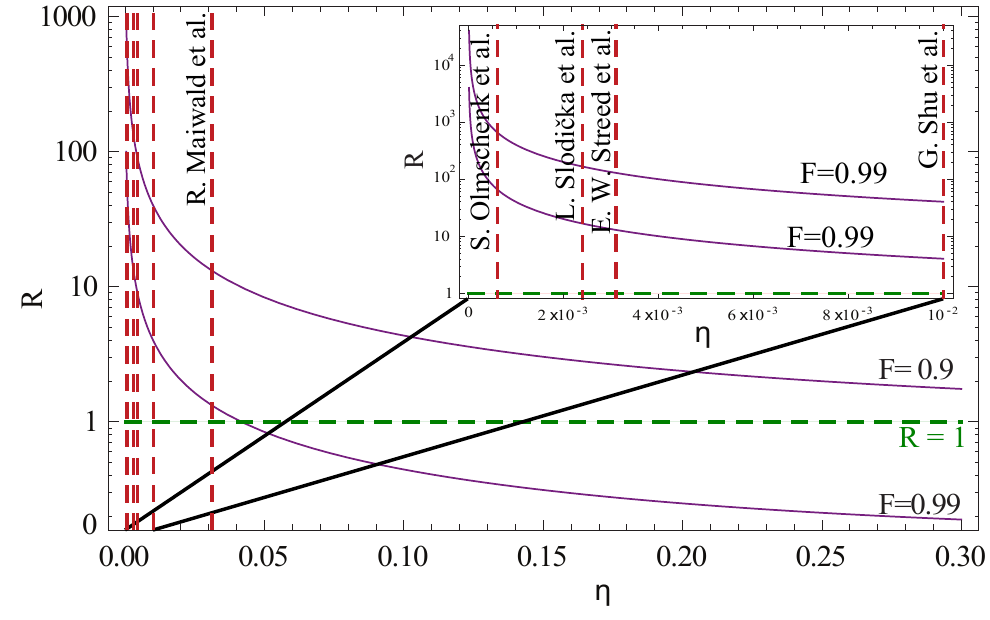}
\caption{Success probability ratio of entanglement generation for the single-photon and two-photon protocols. For current ion-trapping experimental setups \cite{Olm09,Slo13, Str11,Shu11,Mai12} the overall collection and detection efficiency is limited to few percents. For such realistic detection efficiencies, the single-photon entanglement generation scheme has potential to be several orders of magnitude faster than the two-photon scheme. The three highest detection efficiencies values are from experiments where fluorescence was detected directly, without coupling to optical fiber.  \label{Motivation}}
\end{center}
\end{figure}

Following the simplified model in the work of Zippilli et al.~\cite{Zip08}, the fidelity and success rate of the single-photon protocol are given by
\begin{equation}
F_1 \sim (1 -p_{\rm e})/(1 - \eta p_{\rm e})\,\,\,\,\,{\rm and}\,\,\,\,\,P_{\rm s,1} \sim 2 \eta p_{\rm e} (1 - \eta p_{\rm e}).
\label{effSingle}
\end{equation}
Here $p_{\rm e}$ is the probability of the successful excitation and emission of a single photon by a single ion. For a given value of $p_{\rm e}$, the fidelity increases with overall detection efficiency because the likelihood of detecting events where two photons are scattered increases. For a two-photon protocol, the effect of detection efficiency on generated state fidelity is negligible, because both atoms need to be excited and only coincidence detection events trigger entanglement, and thus the fidelity of the generated state with the maximally entangled state is assumed to be $F_2=1$. However, the rate and \keyword{success probability} of entanglement generation depend here quadratically on $\eta$,
\begin{equation}
P_{\rm s,2} \sim \eta^2,
\end{equation}
since the two photons need to be detected at the same time.

Fig.~\ref{Motivation} shows the ratio of success probabilities $R=P_{\rm s,1}/P_{\rm s,2}$ of the two protocols for fixed values of the generated states fidelities as a function of detection efficiency. For a given desired fidelity the two-photon scheme is faster only for high overall detection efficiencies. There is a large advantage in using the single-photon scheme for experimental setups with detection efficiencies below $10^{-2}$. For most of currently realized single-atom experiments, the theoretical gain in entanglement generation rate using the single-photon scheme thus corresponds to several orders of magnitude. In addition, even for unrealistically high detection efficiencies of more than 90\,\%, the single-photon scheme can give higher success rates of generated entangled states with high fidelities. This is due to the high detection probability of double excitations in this limit, which correspond to the fundamental source of infidelity in the single-photon protocol.

\subsection{Generation of entanglement by a single photon detection}

The entanglement of distant single atoms through the detection of a single photon was proposed in the seminal work of Cabrillo et al.~\cite{Cab99}. In this scheme, two atoms (A,B) are initially both prepared in the same long-lived electronic state $|gg\rangle$, see Fig.~\ref{fig:protocolsEnt}. Each atom is then excited with a small probability $p_e$ to another long-lived state $|e\rangle$ through a spontaneous \keyword{Raman process} ($|g\rangle\rightarrow|i\rangle\rightarrow|e\rangle $) by weak excitation of the $|g\rangle\rightarrow|i\rangle$ transition and spontaneous emission of the single photon on the $|i\rangle\rightarrow|e\rangle$ transition. Here $|i\rangle$ denotes an auxiliary atomic state with short lifetime. This Raman process entangles each of the atom's internal states $|s \rangle$ with the emitted photon number $|n \rangle$, so the state of each atom and its corresponding light mode can be written as
\begin{equation}
|s, n\rangle = \sqrt{1-p_e}|g,0\rangle e^{i \phi_{L}} +\sqrt{p_e}|e,1\rangle e^{i \phi_D}.
\end{equation}
The phases $\phi_L$ and $\phi_D$ correspond to the phase of the exciting laser at the position of the atoms and the phase acquired by the spontaneously emitted photons on their way to the detectors, respectively. The total state of the system consisting of both atoms and the light modes can be written as
\begin{equation}
\begin{split}
&|s_{\rm A},s_{\rm B}, n_{\rm A},n_{\rm B}\rangle = (1-p_e)e^{i(\phi_{\rm L,A}+\phi_{\rm L,B})} |gg,00\rangle + \\
& \hspace{+2.8cm} +\sqrt{p_e(1-p_e)}(e^{i(\phi_{\rm L,A}+\phi_{\rm D,B})}|eg,10\rangle+e^{i(\phi_{\rm L,B}+\phi_{\rm D,A})} |ge,01\rangle) + \\
& \hspace{+2.8cm} + p_e e^{i (\phi_{\rm D,A}+\phi_{\rm D,B})}|ee,11\rangle.
\end{split}
\end{equation}

Indistinguishability of the photons from the two atoms is achieved by overlapping their corresponding modes, for example using a beam splitter. Single photon detection then projects the two-atom state onto the entangled state
\begin{equation}
|\Psi^\phi\rangle=\frac{1}{\sqrt{2}}(|eg\rangle+e^{i\phi}|ge\rangle),
\label{stateCreated}
\end{equation}
with the probability of 1-$p_e^2$, where $p_e^2$ is the probability of simultaneous excitation of both atoms. The phase of the generated entangled state $\phi$ corresponds to the sum of the phase differences acquired by the exciting beam at the position of the two atoms and the phase difference acquired by the photons from the respective atoms upon traveling to the detector,
\begin{equation}
\phi=(\phi_{\rm L,B}-\phi_{\rm L,A})+(\phi_{\rm D,A}-\phi_{\rm D,B})).
\label{phi}
\end{equation}

The only limiting factor for the fidelity of the generated state with respect to the maximally entangled state emerging from the presented simplified model is the probability of simultaneous excitation of the two atoms $p_e^2$. However, this can be made arbitrarily small at the expense of entanglement generation success probability $P_{\rm s}$, as demonstrated in Fig.~\ref{Motivation}. The phase of the generated state depends on the relative length of the excitation and detection paths, which therefore need to be stabilized with sub-wavelength precision. Random changes of these path-lengths caused by atomic motion or air density fluctuations change the phase of the entangled state in Eq.~(\ref{stateCreated}) in an incoherent way, which can considerably reduce the fidelity of the generated state. In the experiment we stabilize the phase $\phi$ with interferometric methods to $\phi=0$. The heralded detection of a single photon should then generate the maximally entangled target state
\begin{equation}
|\Psi^+\rangle=\frac{1}{\sqrt{2}}(|eg\rangle+|ge\rangle).
\label{stateCreated}
\end{equation}

\subsection{Experimental realization}

\begin{figure}[t]
\begin{center}
\includegraphics[scale=0.102]{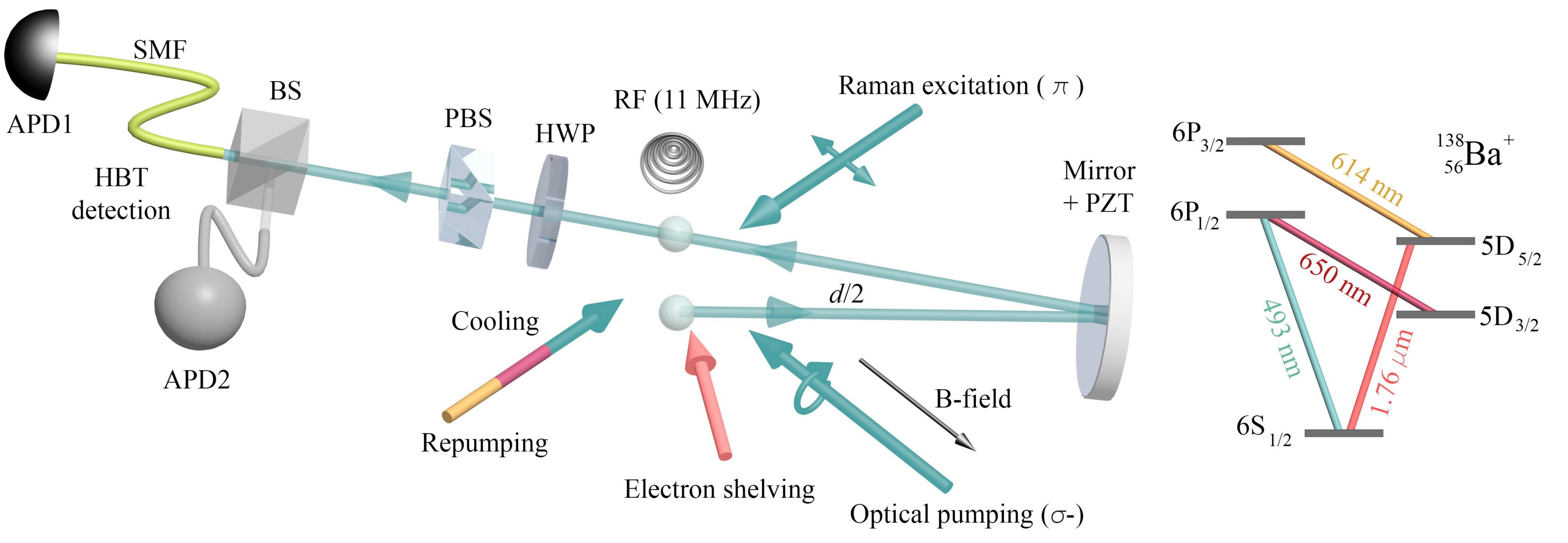}
\caption{Scheme of the experimental setup for entanglement generation by a single photon detection and relevant electronic level scheme of $^{138}{\rm Ba}^+$~\cite{Slo13}. Fluorescence of the two ions is overlapped using a distant mirror which sets the effective distance between them to $d=1$\,meter. A half wave plate (HWP), a polarizing beam splitter (PBS) and a single-mode optical fiber select the polarization and the spatial mode before an avalanche photodiode (APD1). A non-polarizing beam-splitter and an additional avalanche photodiode (APD2) can be inserted to form a Hanbury-Brown-Twiss setup.
\label{setupEnt}}
\end{center}
\end{figure}
For the experimental realization of the single-photon entanglement generation\linebreak scheme two barium ions are trapped in a linear Paul trap setup. As shown in Fig.~\ref{setupEnt}, laser light at 493\,nm is used to Doppler-cool the ions and to detect their electronic states by means of electron shelving, and a laser field at 650\,nm pumps the atoms back to the 6P$_{1/2}$ level from the metastable 5D$_{3/2}$ state. By carefully adjusting the cooling and trapping parameters, the ions are always well within the Lamb-Dicke limit so that the photon \keyword{recoil} during the Raman scattering process is mostly carried by the trap. This ensures that only minimal information is retained in the motion of the ion about which atom has scattered the photon during the entanglement generation process. The fluorescence photons are efficiently collected by two high numerical aperture lenses (NA $\approx 0.4$) placed 14\,mm away from the atoms. A magnetic field of 0.41\,mT is applied at an angle of 40\,degrees with respect to the two-ion axis and defines the quantization axis. After passing through a polarizing beam splitter that blocks the $\pi$-polarized light and lets $\sigma$-polarized light pass, the spatial overlap of the photons is guaranteed by collecting the atomic fluorescence of the first ion in a single mode optical fiber, whilst the fluorescence of the second ion is sent to a distant mirror that retro-reflects it in the same optical fiber. The fluorescence of the two ions (including the Raman scattered light) is then detected by an avalanche photodiode with a quantum efficiency of 60\,\%.

In order to produce a pure entangled state of two qubits, the phase $\phi$ of the generated state, defined in Eq.~(\ref{phi}), must be controlled with high precision. This is achieved by a measurement of the phase of the interference produced by the elastic scattering of the 493\,nm Doppler-cooling beam from the two ions. Scattered photons will follow the same optical paths as the photons scattered by the Raman beam just in opposite directions. Observation of their interference can be then used for stabilization of the relative phase of the exciting Raman beam at the position of the two ions.

\begin{figure}
\begin{center}
\includegraphics[scale=0.13]{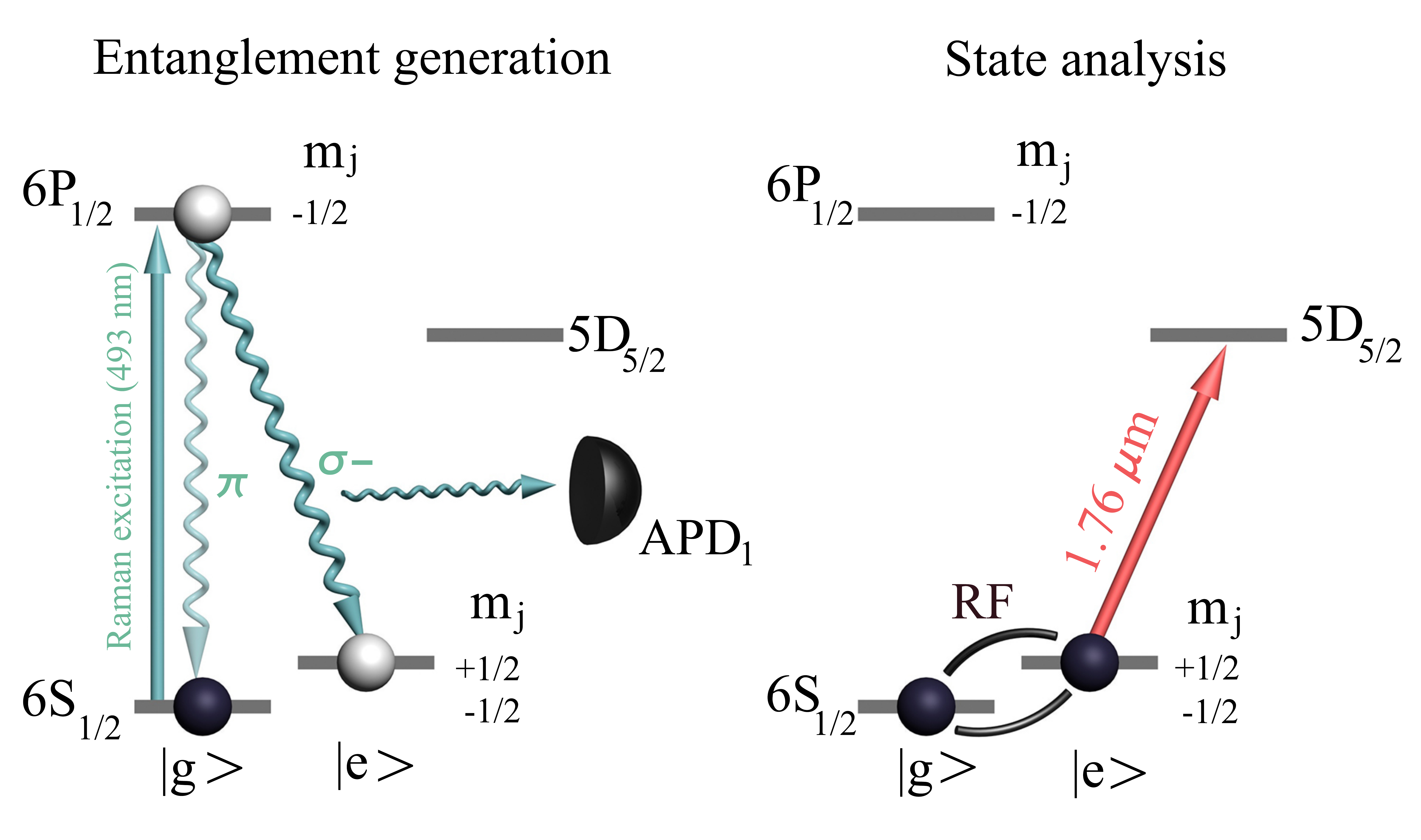}
\caption{Experimental sequence~\cite{Slo13}. Spontaneous Raman scattering from $|g\rangle$ to $|e\rangle$ triggers emission of a single photon from the two atoms. Upon successful detection of a $\sigma^-$ photon, state analysis comprising coherent radio-frequency (RF) pulses at 11.5\,MHz, and electron shelving to the 5D$_{5/2}$ level are performed. \label{sequenceCab}}
\end{center}
\end{figure}

Every experimental sequence of this measurement starts by Doppler-cooling of the two ions. Then the ion-mirror distance $d/2$ is stabilized by locking the position of the measured interference fringe to a chosen position. The electronic states of the ions are then prepared to the Zeeman substate $|6{\rm S}_{1/2, (m=-1/2)}\rangle = |g\rangle$ by optical pumping with a circularly polarized 493\,nm laser pulse propagating along the magnetic field. Next, a weak horizontally polarized laser pulse excites both ions on the S$_{1/2}\leftrightarrow$ P$_{1/2}$ transition with a probability $p_e=0.07$. From the excited state the ion can decay to the other Zeeman sublevel $|6{\rm S}_{1/2,(m=+1/2)}\rangle = |e\rangle$, see Fig.~\ref{sequenceCab}. The electronic state of each ion is at this point entangled with the number of photons $|0\rangle$ or $|1\rangle$ in the $\sigma^-$ polarized photonic mode. Provided that high indistinguishability of the two photonic channels is assured and that simultaneous excitation of both atoms is negligible, detection of a single $\sigma^-$ photon on the APD projects the two-ion state onto the maximally entangled state given by Eq.~(\ref{stateCreated}).

Following the detection of a Raman scattered $\sigma^-$ photon, the two-atom state is coherently manipulated to allow for measurements in a different basis used for the estimation of the generated state. As shown in Fig.~\ref{sequenceCab}, this is done by first applying radio-frequency (RF) pulses that are resonant with the $|g\rangle \leftrightarrow |e\rangle$ transition of both atoms at transition frequency of 11.5\,MHz. Finally, discrimination between the two Zeeman sub-levels of the 6S$_{1/2}$ state is done by shelving the population of the 6S$_{1/2,(m=-1/2)}$ state to the metastable 5D$_{5/2}$ level using a narrowband 1.76\,$\mu$m laser. The fluorescence rate on the 6S$_{1/2}\leftrightarrow {\rm 6P}_{1/2}$ transition \cite{Slo13} allows distinguishing between having no excitation at all $\rho_{gg}$, a single delocalized excitation $\rho_{ge}$ or $\rho_{eg}$, and two excitations $\rho_{ee}$ in the two-atoms system. These events can be separated with 98\,\% probability, which enables efficient reconstruction of the relevant parts of the two-atom density matrix. The 614\,nm laser field then resets the ions to the 6S$_{1/2}$ state and the same experiment is repeated 100 times.

\subsubsection{Estimation of the generated state}

The success rate and fidelity of the generated entangled state of
distant ions can be estimated by measuring the overlap of the
generated state with the desired entangled state every time the
heralding photon is detected. It is sufficient to measure only
certain parts of the density matrix which contribute to this
overlap~\cite{Slo13}. The fidelity $F=\langle
\Psi^+|\rho|\Psi^+\rangle$ of the general two-qubit state $\rho$
with the desired maximally entangled state $|\Psi^+\rangle$ reads
\begin{equation}
F=\frac{1}{2} [\rho_{ge}+\rho_{eg}+ 2 {\rm Re}(\rho_{eg,ge})].
\label{fidelityRho}
\end{equation}
The fidelity thus depends only on the sum of diagonal populations $\rho_{ge}$ and $\rho_{eg}$ and on the real part of the off-diagonal term $\rho_{eg,ge}$ that expresses the mutual coherence between them. All these terms can be accessed using the collective rotations
\begin{equation}
\hat{R}(\theta,\phi)=\exp{\left[-i\frac{\theta}{2}\left(\cos\phi\hat{S}_x+\sin\phi\hat{S}_y\right)\right]},
\end{equation}
followed by the measurement of parity operator
\begin{equation}
\hat{P}=\hat{p}_{gg} +\hat{p}_{ee}-\hat{p}_{eg}-\hat{p}_{ge},
\end{equation}
where $\hat{p}_{ij}$ are the projection operators on states $|ij\rangle$, $i,j\in \{g,e\}$ in different bases~\cite{Sac00} and $\hat{S}_{x,y}=\hat{\sigma}_{x,y}^{(1)}\otimes I^{(2)}+I^{(1)}\otimes \hat{\sigma}_{x,y}^{(2)}$ is the total angular momentum operator in x- or y-direction for both ions. The rotation angle $\theta$ and rotation axis $\phi$ on the Bloch sphere are determined by the duration and the phase of the RF pulses, respectively.

For the state $|\Psi^+\rangle$, it can be readily shown that
\begin{equation}
{\rm Tr}\left[\hat{P}\hat{R}(\pi/2,\phi)|\Psi^+\rangle\langle\Psi^+|\hat{R}^\dagger (\pi/2,\phi)\right]=1
\end{equation}
for all $\phi$. A parity measurement on the $|\Psi^+\rangle$
entangled state is therefore invariant with respect to the change
of the rotation pulse $\hat{R}(\pi/2,\phi)$ phase $\phi$. In order
to measure the parity oscillations for this state, it first has to
be rotated by a global $\hat{R}(\pi/2,\pi/2)$ pulse, corresponding
to a $\hat{\sigma}_y$ rotation on both qubits with the pulse area
of $\pi/2$.

\subsubsection{Entanglement generation results}

\begin{figure}[!t!]
\begin{center}
\includegraphics[scale=2.0]{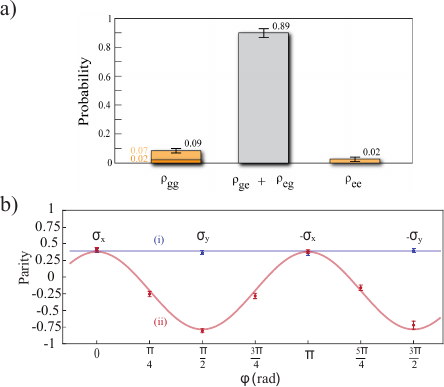}
\caption{Characterization of the entangled state~\cite{Slo13}. a)
Two-atom state populations after the detection of a $\sigma^-$
photon showing that the total probability of measuring the state
with a single excitation is (89$\pm 3$)\,\%. Spurious populations
of the $|gg\rangle$ state are caused by double excitations of each
ion (0.07) and dark count rate of the employed avalanche diode
(0.02). State $|ee\rangle$ is populated due to the simultaneous
excitation of the two ions. b) Parity measurements as a function
of the RF-phase. Trace (ii) corresponds to the measurement of the
atomic populations after two global rotations
$\hat{R}(\pi/2,\pi/2)\hat{R}(\pi/2,\phi)$. In the measurement of
trace (i) only a single global RF-pulse $\hat{R}(\pi/2,\phi)$ is
applied.\label{Parity}}
\end{center}
\end{figure}

The electronic state of two ions is analyzed after each heralding
photon detection. Fig.~\ref{Parity}-a) shows that in (89$\pm
3$)\,\% of the heralded events correspond to the events where only
one of the atoms was excited to the $|e\rangle$ state. This is in
good agreement with the excitation probability $p_{\rm e}=0.07 \pm
0.03$ of each ion and the measured dark-count rate of the employed
avalanche photodiode of 10\,counts/s.

Fig.~\ref{Parity}-b), trace (ii), shows the results of the parity
operator $\hat{P}$ measurements that are, as explained above,
preceded by two global RF rotations
$\hat{R}(\pi/2,\pi/2)\hat{R}(\pi/2,\phi)$ for estimation of the
quantum coherence of the generated state. The first applied pulse
$\hat{R}(\pi/2,\pi/2)$ performs the unitary rotation
$\hat{R}(\pi/2,\pi/2)|\Psi^+\rangle \rightarrow |\Phi^-\rangle$,
where $|\Phi^-\rangle=\frac{1}{\sqrt{2}}(|gg\rangle-|ee\rangle)$.
The second RF-pulse with same duration but with a phase $\phi$
then performs the rotation $\hat{R}(\pi/2,\phi)|\Phi^-\rangle$.
The measured parity signal clearly oscillates as a function of
phase $\phi$ with contrast of (58.0$\pm 2.5$)\,\% and period of
$\pi$, a proof that we indeed succeed in preparing an entangled
two-ion state close to $|\Psi^+\rangle$~\cite{Sac00}. The mean
value of the parity operator at zero phase
$\langle\hat{P}\rangle_{\phi\rightarrow 0}$ corresponds to the
difference between the inner parts and outer-most coherence terms
of the density matrix. The measured value corresponds to
$2\rm{Re}(\rho_{ge,eg}-\rho_{gg,ee})=0.38\pm 0.03$. To precisely
quantify the fidelity of the generated state with
$|\Psi^+\rangle$, the real part of the coherence term
$\rho_{ge,eg}$ needs to be estimated. This is done by measuring
the parity without the first RF rotation. Trace (i) of
Fig.~\ref{Parity}-b) shows the expectation value of the parity as
a function of the phase $\phi$ of the single RF-pulse. The
invariance of the measurement result with respect to the phase
$\phi$ proves that $\rho_{gg,ee}=0\pm 0.03$, so that indeed only
the coherence corresponding to the state $|\Psi^+\rangle$ is
measured. The estimated fidelity of the generated state with the
maximally entangled state is $|\Psi^+\rangle$ is $F=(63.5\pm
2)\,\%$. The threshold for an entanglement is thus surpassed by
more than six standard deviations. The coherence between the
$|ge\rangle$ and $|eg\rangle$ states of $(38\pm 3)\,\%$ is limited
by three main processes ~\cite{Slo13}. First, imperfect
populations of $|ge\rangle$ and $|eg\rangle$ states set a limit of
89\,\%~\cite{Shi06}. Around 4\,\% of the coherence loss can be
attributed to the finite coherence time of the individual atomic
qubits (120\,$\mu$s) due to collective magnetic field
fluctuations. Although the generated $|\Psi^+\rangle$ state is
intrinsically insensitive against collective dephasing~\cite{Kie01,Roo04},
a loss of coherence is indeed expected after a
rotation of $|\Psi^+\rangle$ out of the decoherence-free subspace.
The highest contribution to the coherence loss can be attributed
to atomic motion, which can provide information about which atom
emitted the photon. Around 55\,\% of the coherence is lost due to
the atomic recoil kicks during the Raman scattering.

The overall fidelity of the maximally entangled state
$|\Psi^+\rangle$ with the experimentally generated one is limited
mainly by the imperfect populations of the desired $|ge \rangle$
and $|eg\rangle$ states and coherence loss due to the atomic
recoil kicks during the Raman scattering. The effect of
motion-induced decoherence can be reduced by cooling the radial
modes to the motional ground state~\cite{Slo12} or by choosing a
forward Raman scattering scenario~\cite{Cab99}. Error bars in the
presented measurements results correspond to one standard
deviation and are estimated statistically from several
experimental runs each giving approximately 120 measurement
outcomes. Up to 60\,\% of the measurement error is caused by the
quantum projection noise. Additional uncertainty comes from slow
magnetic field drift with a magnitude of several tens of nT making
the RF-driving off-resonant by tens of kHz.


An important advantage of the single-photon heralding mechanism is
the possibility of achieving a high entanglement generation rate.
In the presented experiment, the entanglement generation rate has
reached (14.0$\pm 1.5$)\,events/minute with an experimental duty
cycle of 2.3\,kHz~\cite{Slo13}. The probability of successful
entanglement generation per each experimental trial estimated from
the single photon detection efficiency and the measured
probability of Raman scattering is $P_{\rm succ}=2 p_e \eta =
1.1\times 10^{-4}$ that corresponds to 15.4 successful
entanglement generation events/minute. The factor two here
corresponds to the probability of detecting a single photon from
one of two ions. The efficiency of detecting a single
Raman-scattered photon was estimated to be $\eta =
8\times\,10^{-4}$. It was derived from the detection probability
of a single Raman scattered photon given by the collection
efficiency of high-NA lenses ($\sim 0.04$), the single-mode fiber
coupling efficiency ($\sim 0.1$) and by the avalanche-photodiode
detection efficiency ($\sim 0.6$). Additional factors of 0.5 and
0.66 come from the polarization filtering of unwanted
$\pi$-polarized photons and from the probability for the ion to
decay back to the $|g\rangle$ state after the Raman pulse
excitation, respectively. The single ion excitation probability
was $p_e = (0.07\pm 0.03)$\,\%. For comparison, the two-photon
heralding entanglement scheme proposed by Simon et
al.~\cite{Sim03} would  for the employed experimental setup give
approximately $P_{\rm succ} \approx 2 \eta^2 = 1.3\times 10^{-6}$,
so about two orders of magnitude smaller success probability of
entanglement generation. For simplicity, $p_e=1$ was assumed here
for the two-photon scheme and an additional factor of 2 accounted
from the two possible contributions to coincidence detection
events.

\subsection{Summary}

In this chapter we have reviewed two methods for free-space
coupling between single ions and photons: the coupling of a weak
probe laser to a single ion in free space (section~\ref{sec:1}),
and the probabilistic generation of entanglement between two ions
by detecting a single scattered photon (section~\ref{sec:2}). Both
method rely on high-numerical aperture optics in order to achieve
a sufficiently high efficiency.

In the experiments presented in section~\ref{sec:1} we have
investigated the free-space interaction of a single ion to a weak
near-resonant probe field. The coupling mediated by a single
objective covering 4\,\% of the full solid angle resulted in an
extinction of the probe field of 1.5\,\%, a phase shift of up to
0.3~degrees, and the observation of electromagnetically-induced
transparency from single atom. Current experimental efforts to
increase the numerical aperture should significantly improve the
interaction strength (see chapter by Leuchs \& Sondermann), which will likely lead
to a number of direct applications of these effects in the field
of quantum information, quantum feedback or single atom
magnetometry. Utilization of a single ion as an optical mirror in
a Fabry-P\'erot-like cavity set-up enabled the observation of
almost full suppression and enhancement by a factor of two of the
atomic coupling constant in the laser probe mode. Besides the
appealing quantum memory applications of such a 
set-up~\cite{Wan12}, the single ion mirror has the potential to become
useful for the realization of an optical switch similar to the
single atom transistors using EIT. Furthermore, the presented
experiment enables to study the quantum electrodynamics in an
exciting regime where both the free-space coupling of the probe
beam and the modification of the vacuum mode density at the
position of an ion play an important role~\cite{het10}.

In section~\ref{sec:2} of this chapter we have summarized an
experimental realization of a proposal by Cabrillo et
al.~\cite{Cab99} where the detection of a single scattered photon
generates entanglement between two ions. This presents an
important step towards the realization of the quantum information
networks with ions and photons. The maximally entangled state
$|\Psi^+\rangle$ was produced with a fidelity of 63.5\,\% and with
entanglement generation rates of 14\,events/minute, which is more
than two orders of magnitude higher than the rate obtainable with
protocols relying on a two-photon coincidence events with the
presented experimental parameters. These results can be further
improved by cooling all of the involved motional modes close to
their ground state~\cite{Slo12} or by choosing a different
excitation direction to minimize residual which-way information.

There are some obvious questions regarding the technical difficulties related to phase stability requirements and photon recoil problems of the single-photon entanglement generation scheme. For generation of the entanglement between distant atoms, the paths of the excitation and detection channels need to be interferometrically stable. This issue has been addressed by the community developing fiber links for comparing remote optical clocks. Recently, coherent laser light transfer over more than 900 km has been shown with a precision exceeding the requirements of the single-photon protocol~\cite{Pre12}. The problem of a which-way information available due to the atomic recoil upon scattering of single photon can be eliminated by changing the geometry of the system. These improvements, together with the experimental results presented, have potential to enable efficient creation and distribution of entanglement between distant sites with well-defined and controllable atomic qubits.

\begin{acknowledgement} {
We would like to thank all our colleagues who were involved in this work over the course of the years, in particular Nadia R\"ock, Philipp Schindler, Daniel Higginbottom, Fran\c{c}ois Dubin, Alexander Gl\"atzle, Muir Kumph, Gabriel Araneda, Sebastian Gerber, Daniel Rotter, Pavel Bushev, Yves Colombe, and J\"urgen Eschner.
The work reported in this chapter has been supported by the Austrian Science Fund FWF (SINFONIA, SFB FoQuS), by the European Union (CRYTERION), by the Institut f\"ur Quanteninformation GmbH, and a Marie Curie International Incoming Fellowship within the 8th European Framework Program.
The writing of this chapter was supported by the European Research Council project QuaSIRIO.
} \end{acknowledgement}

\bibliographystyle{SpringerPhysMWM} 

\bibliography{Slodicka_bib}

\begin{thebibliography}{10}

\bibitem{Cir97}
Cirac, J.~I., Zoller, P., Kimble, H.~J., and Mabuchi, H.
\newblock Quantum state transfer and entanglement distribution among distant
  nodes in a quantum network.
\newblock {\em Phys. Rev. Lett.}{ \bf 78}, 3221--3224 (1997).

\bibitem{Bri98}
Briegel, H.-J., D\"ur, W., Cirac, J.~I., and Zoller, P.
\newblock Quantum repeaters: The role of imperfect local operations in quantum
  communication.
\newblock {\em Phys. Rev. Lett.}{ \bf 81}, 5932--5935 (1998).

\bibitem{Dua01}
Duan, L., Lukin, M., Cirac, I., and Zoller, P.
\newblock Long-distance quantum communication with atomic ensembles and linear
  optics.
\newblock {\em Nature}{ \bf 414}, 413--418 (2001).

\bibitem{Bru94}
Brune, M. {\em et~al.}
\newblock From {Lamb} shift to light shifts: Vacuum and subphoton cavity fields
  measured by atomic phase sensitive detection.
\newblock {\em Phys. Rev. Lett.}{ \bf 72}, 3339--3342 (1994).

\bibitem{Pin00}
Pinkse, P. W.~H., Fischer, T., Maunz, P., and Rempe, G.
\newblock Trapping an atom with single photons.
\newblock {\em Nature}{ \bf 404}, 365--368 (2000).

\bibitem{Hoo00}
Hood, C.~J., Lynn, T.~W., Doherty, A.~C., Parkins, A.~S., and Kimble, H.~J.
\newblock The atom-cavity microscope: Single atoms bound in orbit by single
  photons.
\newblock {\em Science}{ \bf 287}, 1447--1453 (2000).

\bibitem{Pol04}
Julsgaard, B., Sherson, J., Cirac, J., Fiur{\'a}{\v{s}}ek, J., and Polzik, E.
\newblock Experimental demonstration of quantum memory for light.
\newblock {\em Nature}{ \bf 432}, 482--486 (2004).

\bibitem{Hau99}
Hau, L.~V., Harris, S.~E., Dutton, Z., and Behroozi, C.~H.
\newblock Light speed reduction to 17 metres per second in an ultracold atomic
  gas.
\newblock {\em Nature}{ \bf 397}, 594--598 (1999).

\bibitem{Phi01}
Phillips, D., Fleischhauer, A., Mair, A., Walsworth, R., and Lukin, M.
\newblock Storage of light in atomic vapor.
\newblock {\em Phys. Rev. Lett.}{ \bf 86}, 783--786 (2001).

\bibitem{Sor07}
Sortais, Y. R.~P. {\em et~al.}
\newblock Diffraction-limited optics for single-atom manipulation.
\newblock {\em Phys. Rev. A}{ \bf 75}, 013406 (2007).

\bibitem{Son07}
Sondermann, M. {\em et~al.}
\newblock Design of a mode converter for efficient light-atom coupling in free
  space.
\newblock {\em Applied Physics B: Lasers and Optics}{ \bf 89}, 489--492 (2007).

\bibitem{Tey08}
Tey, M. {\em et~al.}
\newblock Strong interaction between light and a single trapped atom without
  the need for a cavity.
\newblock {\em Nature Physics}{ \bf 4}, 924--927 (2008).

\bibitem{Zum08}
Zumofen, G., Mojarad, N., Sandoghdar, V., and Agio, M.
\newblock Perfect reflection of light by an oscillating dipole.
\newblock {\em Phys. Rev. Lett.}{ \bf 101}, 180404 (2008).

\bibitem{Ger07}
Gerhardt, I. {\em et~al.}
\newblock Strong extinction of a laser beam by a single molecule.
\newblock {\em Phys. Rev. Lett.}{ \bf 98}, 33601 (2007).

\bibitem{Wri08}
Wrigge, G., Gerhardt, I., Hwang, J., Zumofen, G., and Sandoghdar, V.
\newblock Efficient coupling of photons to a single molecule and the
  observation of its resonance fluorescence.
\newblock {\em Nat. Phys.}{ \bf 4}, 60--66 (2007).

\bibitem{Vam07}
Vamivakas, A.~N. {\em et~al.}
\newblock Strong extinction of a far-field laser beam by a single quantum dot.
\newblock {\em Nano letters}{ \bf 7}, 2892--2896 (2007).

\bibitem{Alj09}
Aljunid, S.~A. {\em et~al.}
\newblock Phase shift of a weak coherent beam induced by a single atom.
\newblock {\em Phys. Rev. Lett.}{ \bf 103}, 153601 (2009).

\bibitem{Hwa09}
Hwang, J. {\em et~al.}
\newblock A single-molecule optical transistor.
\newblock {\em Nature}{ \bf 460}, 76--80 (2009).

\bibitem{Slo10}
Slodi\ifmmode~\check{c}\else \v{c}\fi{}ka, L., H\'etet, G., Gerber, S.,
  Hennrich, M., and Blatt, R.
\newblock Electromagnetically induced transparency from a single atom in free
  space.
\newblock {\em Phys. Rev. Lett.}{ \bf 105}, 153604 (2010).

\bibitem{Het11}
H\'etet, G., Slodi\ifmmode~\check{c}\else \v{c}\fi{}ka, L., Hennrich, M., and
  Blatt, R.
\newblock Single atom as a mirror of an optical cavity.
\newblock {\em Phys. Rev. Lett.}{ \bf 107}, 133002 (2011).

\bibitem{Het13}
H{\'e}tet, G., Slodi{\v{c}}ka, L., R{\"o}ck, N., and Blatt, R.
\newblock Free-space read-out and control of single-ion dispersion using
  quantum interference.
\newblock {\em Physical Review A}{ \bf 88}, 041804 (2013).

\bibitem{Fle05}
Fleischhauer, M., Imamo\u{g}lu, A., and Marangos, J.~P.
\newblock Electromagnetically induced transparency: Optics in coherent media.
\newblock {\em Rev. Mod. Phys.}{ \bf 77}, 633--673 (2005).

\bibitem{Ber98a}
Bergmann, K., Theuer, H., and Shore, B.
\newblock Coherent population transfer among quantum states of atoms and
  molecules.
\newblock {\em Rev. Mod. Phys.}{ \bf 70}, 1003--1025 (1998).

\bibitem{Eis05}
Eisaman, M.~D. {\em et~al.}
\newblock Electromagnetically induced transparency with tunable single-photon
  pulses.
\newblock {\em Nature}{ \bf 438}, 837--841 (2005).

\bibitem{Boo07}
Boozer, A.~D., Boca, A., Miller, R., Northup, T.~E., and Kimble, H.~J.
\newblock Reversible state transfer between light and a single trapped atom.
\newblock {\em Phys. Rev. Lett.}{ \bf 98}, 193601 (2007).

\bibitem{Mue10}
M{\"u}cke, M. {\em et~al.}
\newblock Electromagnetically induced transparency with single atoms in a
  cavity.
\newblock {\em Nature}{ \bf 465}, 755--758 (2010).

\bibitem{Rit12}
Ritter, S. {\em et~al.}
\newblock An elementary quantum network of single atoms in optical cavities.
\newblock {\em Nature}{ \bf 484}, 195--200 (2012).

\bibitem{Hae08}
H\"affner, H., Roos, C., and Blatt, R.
\newblock Quantum computing with trapped ions.
\newblock {\em Physics Reports}{ \bf 469}, 155 (2008).

\bibitem{Mau07}
Maunz, P. {\em et~al.}
\newblock Quantum interference of photon pairs from two remote trapped atomic
  ions.
\newblock {\em Nat Phys}{ \bf 3}, 538–541 (2007).

\bibitem{Ger09njp}
Gerber, S. {\em et~al.}
\newblock Quantum interference from remotely trapped ions.
\newblock {\em New Journal of Physics}{ \bf 11}, 013032 (2009).

\bibitem{Koc94}
Kochan, P. and Carmichael, H.~J.
\newblock Photon-statistics dependence of single-atom absorption.
\newblock {\em Phys. Rev. A}{ \bf 50}, 1700--1709 (1994).

\bibitem{Tey09}
Tey, M.~K. {\em et~al.}
\newblock Interfacing light and single atoms with a lens.
\newblock {\em New Journal of Physics}{ \bf 11}, 043011 (2009).

\bibitem{Venk04}
van Enk, S.~J.
\newblock Atoms, dipole waves, and strongly focused light beams.
\newblock {\em Phys. Rev. A}{ \bf 69}, 043813 (2004).

\bibitem{Esc01}
Eschner, J., Raab, C., Schmidt-Kaler, F., and Blatt, R.
\newblock Light interference from single atoms and their mirror images.
\newblock {\em Nature}{ \bf 413}, 495--498 (2001).

\bibitem{Ima89}
Imamo\u{g}lu, A. and Harris, S.~E.
\newblock Lasers without inversion: interference of dressed lifetime-broadened
  states.
\newblock {\em Opt. Lett.}{ \bf 14}, 1344--1346 (1989).

\bibitem{Rabl}
Rabl, P., Steixner, V., and Zoller, P.
\newblock Quantum-limited velocity readout and quantum feedback cooling of a
  trapped ion via electromagnetically induced transparency.
\newblock {\em Phys. Rev. A}{ \bf 72}, 043823 (2005).

\bibitem{het10}
H\'etet, G., Slodi\ifmmode~\check{c}\else \v{c}\fi{}ka, L., Gl\"atzle, A.,
  Hennrich, M., and Blatt, R.
\newblock {\textsc{QED}} with a spherical mirror.
\newblock {\em Phys. Rev. A}{ \bf 82}, 063812 (2010).

\bibitem{Jia07}
Jiang, L., Taylor, J.~M., S{\o}rensen, A.~S., and Lukin, M.~D.
\newblock Distributed quantum computation based on small quantum registers.
\newblock {\em Phys. Rev. A}{ \bf 76}, 062323 (2007).

\bibitem{Cir99}
Cirac, J.~I., Ekert, A.~K., Huelga, S.~F., and Macchiavello, C.
\newblock Distributed quantum computation over noisy channels.
\newblock {\em Phys. Rev. A}{ \bf 59}, 4249 (1999).

\bibitem{Got99}
Gottesman, D. and Chuang, I.~L.
\newblock Demonstrating the viability of universal quantum computation using
  teleportation and single-qubit operations.
\newblock {\em Nature}{ \bf 402}, 390--393 (1999).

\bibitem{Dua10}
Duan, L.~M. and Monroe, C.
\newblock Colloquium: Quantum networks with trapped ions.
\newblock {\em Rev. Mod. Phys.}{ \bf 82}, 1209 (2010).

\bibitem{Cab99}
Cabrillo, C., Cirac, J.~I., Garcia-Fern\'andez, P., and Zoller, P.
\newblock Creation of entangled states of distant atoms by interference.
\newblock {\em Phys. Rev. A}{ \bf 59}, 1025--1033 (1999).

\bibitem{Sim03}
Simon, C. and Irvine, W. T.~M.
\newblock Robust long-distance entanglement and a loophole-free {Bell} test
  with ions and photons.
\newblock {\em Phys. Rev. Lett.}{ \bf 91}, 110405 (2003).

\bibitem{Cho05}
Chou, C. {\em et~al.}
\newblock Measurement-induced entanglement for excitation stored in remote
  atomic ensembles.
\newblock {\em Nature}{ \bf 438}, 828--832 (2005).

\bibitem{Cho07}
Chou, C. {\em et~al.}
\newblock Functional quantum nodes for entanglement distribution over scalable
  quantum networks.
\newblock {\em Science}{ \bf 316}, 1316--1320 (2007).

\bibitem{Moe07}
Moehring, D.~L. {\em et~al.}
\newblock Entanglement of single-atom quantum bits at a distance.
\newblock {\em Nature}{ \bf 449}, 68--71 (2007).

\bibitem{Zip08}
Zippilli, S. {\em et~al.}
\newblock Entanglement of distant atoms by projective measurement: the role of
  detection efficiency.
\newblock {\em New Journal of Physics}{ \bf 10}, 103003 (2008).

\bibitem{Luo09}
Luo, L. {\em et~al.}
\newblock Protocols and techniques for a scalable atom--photon quantum network.
\newblock {\em Fortschritte der Physik}{ \bf 57}, 1133--1152 (2009).

\bibitem{Rau01}
Raussendorf, R. and Briegel, H.~J.
\newblock A one-way quantum computer.
\newblock {\em Physical Review Letters}{ \bf 86}, 5188–5191 (2001).

\bibitem{Dua05}
Duan, L.-M. and Raussendorf, R.
\newblock Efficient quantum computation with probabilistic quantum gates.
\newblock {\em Physical Review Letters}{ \bf 95} (2005).

\bibitem{Str11}
Streed, E.~W., Norton, B.~G., Jechow, A., Weinhold, T.~J., and Kielpinski, D.
\newblock Imaging of trapped ions with a microfabricated optic for quantum
  information processing.
\newblock {\em Phys. Rev. Lett.}{ \bf 106}, 010502 (2011).

\bibitem{Str12}
Streed, E., Jechow, A., Norton, B., and Kielpinski, D.
\newblock Absorption imaging of a single atom.
\newblock {\em Nature Communications}{ \bf 3}, 933 (2012).

\bibitem{Dar05}
Darqui\'e, B. {\em et~al.}
\newblock Controlled single-photon emission from a single trapped two-level
  atom.
\newblock {\em Science}{ \bf 309}, 454--456 (2005).

\bibitem{Olm10}
Olmschenk, S. {\em et~al.}
\newblock Quantum logic between distant trapped ions.
\newblock {\em International Journal of Quantum Information}{ \bf 8}, 337--394
  (2010).

\bibitem{Shu11}
Shu, G., Chou, C.~K., Kurz, N., Dietrich, M., and Blinov, B.
\newblock Efficient fluorescence collection and ion imaging with the "tack" ion
  trap.
\newblock {\em JOSA B}{ \bf 28}, 2865--2870 (2011).

\bibitem{Mai09}
Maiwald, R. {\em et~al.}
\newblock Stylus ion trap for enhanced access and sensing.
\newblock {\em Nature Physics}{ \bf 5}, 551--554 (2009).

\bibitem{Sto09}
Stobi{\'n}ska, M., Alber, G., and Leuchs, G.
\newblock Perfect excitation of a matter qubit by a single photon in free
  space.
\newblock {\em Europhysics Letters}{ \bf 86}, 14007 (2009).

\bibitem{Olm09}
Olmschenk, S. {\em et~al.}
\newblock Quantum teleportation between distant matter qubits.
\newblock {\em Science}{ \bf 323}, 486--489 (2009).

\bibitem{Slo13}
Slodi{\v{c}}ka, L. {\em et~al.}
\newblock Atom-atom entanglement by single-photon detection.
\newblock {\em Physical review letters}{ \bf 110}, 083603 (2013).

\bibitem{Mai12}
Maiwald, R. {\em et~al.}
\newblock Collecting more than half the fluorescence photons from a single ion.
\newblock {\em Phys. Rev. A}{ \bf 86}, 043431 (2012).

\bibitem{Sac00}
Sackett, C.~A. {\em et~al.}
\newblock Experimental entanglement of four particles.
\newblock {\em Nature}{ \bf 404}, 256--259 (2000).

\bibitem{Shi06}
Shirokov, M.~I.
\newblock Cauchy inequality and uncertainty relations for mixed states.
\newblock {\em International Journal of Theoretical Physics}{ \bf 45}, 141--151
  (2006).

\bibitem{Kie01}
Kielpinski, D. {\em et~al.}
\newblock A decoherence-free quantum memory using trapped ions.
\newblock {\em Science}{ \bf 291}, 1013--1015 (2001).

\bibitem{Roo04}
Roos, C.~F. {\em et~al.}
\newblock Bell states of atoms with ultralong lifetimes and their tomographic
  state analysis.
\newblock {\em Phys. Rev. Lett.}{ \bf 92}, 220402 (2004).

\bibitem{Slo12}
Slodi\ifmmode~\check{c}\else \v{c}\fi{}ka, L. {\em et~al.}
\newblock Interferometric thermometry of a single sub-{Doppler}-cooled atom.
\newblock {\em Phys. Rev. A}{ \bf 85}, 043401 (2012).

\bibitem{Wan12}
Wang, Y., Min\'a\ifmmode~\check{r}\else \v{r}\fi{}, J., H\'etet, G., and
  Scarani, V.
\newblock Quantum memory with a single two-level atom in a half cavity.
\newblock {\em Phys. Rev. A}{ \bf 85}, 013823 (2012).

\bibitem{Pre12}
Predehl, K. {\em et~al.}
\newblock A 920-kilometer optical fiber link for frequency metrology at the
  19th decimal place.
\newblock {\em Science}{ \bf 336}, 441--444 (2012).

\end{thebibliography}

\printindex
\end{document}